\documentclass[acp, manuscript]{copernicus}

\usepackage{amsmath}

\usepackage{tikz}
\usetikzlibrary{patterns}
\usetikzlibrary{shapes}
\usetikzlibrary{decorations.text}
\usetikzlibrary{shapes.multipart}
\usetikzlibrary{arrows,shapes.geometric,positioning}
\usetikzlibrary{shapes,positioning,decorations.pathmorphing}
\usetikzlibrary{calc}
\usepackage{booktabs}
\usetikzlibrary{backgrounds}
\usepackage{varwidth}
\usetikzlibrary{fit}

\usepackage{latexsym,euscript,textcomp}
\usepackage{color,graphics,epsf,dcolumn}
\usepackage{epstopdf,ulem}
\usepackage{blindtext}
\usepackage{threeparttable}
\newcommand{\beq}{\begin{equation}}
\newcommand{\eeq}{\end{equation}}
\newcommand{\pt}{\partial}

\begin{document}
\nolinenumbers

\title{On the possible role of condensation-related hydrostatic pressure adjustments in intensification and weakening of tropical cyclones}


\Author[1,2]{Anastassia M.}{Makarieva}
\Author[1]{Andrei V.}{Nefiodov}

\affil[1]{Theoretical Physics Division, Petersburg Nuclear Physics Institute, Gatchina  188300, St.~Petersburg, Russia}
\affil[2]{Institute for Advanced Study, Technical University of Munich, Lichtenbergstrasse 2 a, D-85748 Garching, Germany}


\runningtitle{Comments on {\textquotedblleft}A Physical Model of Tropical Cyclone Central Pressure Filling at Landfall{\textquotedblright}}

\runningauthor{Makarieva and Nefiodov}

\correspondence{A. M. Makarieva (ammakarieva@gmail.com)}

\received{}
\pubdiscuss{} 
\revised{}
\accepted{}
\published{}


\firstpage{1}

\maketitle

\begin{abstract}
\large
It is shown that condensation and precipitation do not disturb hydrostatic equilibrium if the local pressure sink (the condensation rate expressed in pressure units) is proportional to the local pressure, with a proportionality coefficient $k$ that is independent of altitude. In the real atmosphere, however, the condensation rate depends, among other factors, on the vertical velocity, which varies with height. As a result, condensation generally disturbs hydrostatic equilibrium and induces pressure adjustments through air-mass redistribution. We propose that a profile in which $k$ is maximized in the upper atmosphere leads to additional upward motion and cyclone intensification, whereas a maximum closer to the surface induces downward motion and cyclone weakening. The magnitude of both effects is expected to be set by the strength of the precipitation mass sink. Using observational data, we find that the median intensification and weakening rates---$12$ and $8$~hPa~day$^{-1}$, respectively, measured over six-hour intervals in Atlantic tropical cyclones---amount to about three quarters of the maximum concurrent precipitation rate (multiplied by gravity) in the core precipitation region. This implies intensification under conditions of \textit{positive} vertically integrated air convergence, a regime impossible in modeled dry hurricanes, with the negative pressure tendency arising because precipitation exceeds the vertically integrated moisture convergence by absolute magnitude. The implications of these results for recent studies that evaluate tropical cyclone (de-)intensification using mass continuity equations that neglect the precipitation mass sink are discussed.
\end{abstract}


\large
\section{Introduction}
\label{intr}

The drop in central surface pressure is a key parameter of tropical cyclones, closely linked to their intensity (i.e., maximum wind velocity). Historically, theoretical research on storm intensification has focused primarily on describing changes in maximum velocity \citep{montgomery17b}. \citet{sparks22a} made a notable effort to explain the physics of storm intensification by examining the surface pressure tendency.

In a hydrostatic atmosphere, surface pressure reflects the weight of the atmospheric column. Consequently, the surface pressure tendency indicates the rate at which that column{\textquoteright}s mass change. According to the law of mass conservation, the rate of change of air mass in the column is equal to the vertically integrated air convergence  plus evaporation $E$ minus precipitation $P$. This leads to the following equation for surface pressure tendency:
\beq\label{spt}
\frac{\pt p}{\pt t} = - g \int_{0}^{H}  \mathrm{div} (\rho \mathbf{u}) dz + g (E - P) ,
\eeq
as derived, for example, in \citet[][Eq.~(5)]{trenberth91} and \citet[][Eq.~(23)]{wacker2006}. Here, $p$ is surface pressure, $g$ is the acceleration of gravity, $\rho$ is the mass density of air, $\mathbf{u}$ is the horizontal velocity, and $H$ is a height at which $\rho$ can be assumed negligible.

Assuming approximate axial symmetry of tropical cyclones, the relevant quantity becomes the mean surface pressure tendency within a cylinder of radius $r$, which can be expressed using Eq.~\eqref{spt} as:
\beq\label{dpdt1a}
\frac{\pt \overline{p}}{\pt t} = -\frac{2g}{r} \int\limits_{0}^{H} \rho u_r dz - g \overline{P},
\eeq
as in Eq.~(6) of \citet{la04}. Here, $u_r$ denotes the radial velocity, and the averaging is performed over the cylinder{\textquoteright}s base area. 
Evaporation is neglected, as it is generally small in the storm core \citep[see, e.g.,][their Table1]{ar17}.

Using a numerical model, \citet{la04} found that during the intensification phase of Hurricane Lili (2002), the average pressure drop within a radius of $r = 100$~km from the storm center was approximately $\partial \overline{p}/\partial t = -0.5$~hPa~h$^{-1}$. Over the same period, the average pressure-equivalent mass loss due to precipitation was about three times larger, with $-g\overline{P} = -1.5$~hPa~h$^{-1}$. On this basis, \citet{la04} concluded that  {\textquotedblleft}the amount of atmospheric mass removed via precipitation exceeded that needed to explain the model sea level pressure decrease.{\textquotedblright}

In the case of Hurricane Lili (2002), where 
\beq\label{c}
-g\overline{P}= c \frac{\pt \overline{p}}{\pt t}
\eeq
with $c\sim3$, omitting the precipitation term from the pressure tendency Eq.~\eqref{dpdt1a} and diagnosing pressure change  solely from the air convergence term (the first term on the right-hand side of Eq.~\eqref{dpdt1a}) would result in a substantial error:  the predicted pressure tendency would not only be twice as large in absolute magnitude as the observed value but also incorrect in sign (positive). In other words, neglecting precipitation would suggest a rapidly weakening storm instead of a rapidly intensifying one.

We emphasize that $c > 1$ in Eq.~\eqref{c} implies that the storm intensifies with a {\it positive vertically integrated air convergence} (i.e., more air flows laterally into the cylinder than out). This stands in fundamental contrast to dry circulations, where a surface pressure drop can occur only if the vertically integrated air convergence is negative.
From a dynamical perspective, a value of $c>1$ implies that, in the upper troposphere, the combined pressure-gradient and centrifugal forces are insufficient to export from the storm core as much air as is drawn inward by the low-level pressure gradient.

\citet{wacker2006} reported a similar pattern in a modeled cyclone that deepened by about $60$~hPa over 90~hours (a mean rate of $0.7$~hPa~h$^{-1}$), while the mean precipitation rate in the eyewall over six hours was approximately $3.7$~hPa~h$^{-1}$—a value the authors described as {\textquotedblleft}a remarkable pressure drop{\textquotedblright}. This again implies a positive convergence term and confirms that neglecting the precipitation term in the pressure tendency equation can lead to substantial inaccuracies.

\section{The model of Sparks and Toumi}

Without reference to prior literature on the subject, \citet{sparks22a} based their model of storm weakening on a reduced form of Eq.~\eqref{dpdt1a}, in which the precipitation term was omitted. This same formulation was subsequently applied to the case of storm intensification \citep{sparks22b}.

Considering a hydrostatic atmosphere
\beq\label{he}
p = g \int\limits_{0}^{H} \rho dz,
\eeq
\citet[their Eq.~(3)]{sparks22a} introduced the density-weighted column-mean radial wind velocity at radius $r$ as
\beq\label{Ur}
{\mathcal U}_r \equiv \dfrac{\int\limits_{0}^{H} \rho u_r dz}{\int\limits_{0}^{H} \rho dz}.
\eeq

Using Eqs.~\eqref{he} and \eqref{Ur}, Eq.~\eqref{dpdt1a} becomes
\beq\label{dpdt1}
\frac{\pt \overline{p}}{\pt t} = -2p(r) \frac{\mathcal{U}_r(r)}{r} - g\overline{P}.
\eeq

To derive their model for the central pressure tendency $\pt p_c/\pt t$, \citet[][their Eq.~(5)]{sparks22a} neglected the last term in Eq.~\eqref{dpdt1}  
and assumed that at the central limit $r \to 0$  the function $\mathcal{U}_r/r$ can be approximated by its value (the leading term of the Taylor expansion) at the radius of maximum wind $r_{m}$, so that
\beq\label{st}
\frac{\pt p_c}{\pt t} = - 2 \lim_{r\to 0} p(r) \dfrac{\mathcal{U}_r(r)}{r}= 
- 2 p_c \lim_{r\to 0} \dfrac{\mathcal{U}_r(r)}{r} \simeq  - 2 p_c \frac{\chi}{r_{m0}}.
\eeq 
Here $\chi \equiv \mathcal{U}_r(r_{m0})$ is the column-mean radial velocity evaluated at $r_{m}= r_{m0}$ at the initial point of time $(t=0)$.

In the following, we demonstrate using observational data---rather than model output---that neglecting the final term in Eq.~\eqref{dpdt1} is not justified, as precipitation makes a substantial contribution to the pressure tendency in both intensifying and weakening tropical cyclones.

To analyze the relationship between intensification rate and precipitation, we followed \citet{ar17}. We used the EBTRK dataset (released 27 July 2016; \citealp{demuth06}) together with the 3-hourly TRMM 3B42 precipitation product (version~7). EBTRK records are available at 6-hour intervals.

For 1998--2015, and for each $k$th EBTRK record (with the $(k+1)$th record referring to the same storm), we defined the intensification rate as $I_k \equiv -4(p_{k+1}-p_k)$ (hPa~day$^{-1}$), where $p$ is the minimum central pressure (hPa). We selected tropical storms over land, identified by a negative distance-to-land value in the final EBTRK column, for which the radius of maximum wind is available. This yielded 360 values of the intensification rate $I$: 45 equal to zero, 277 negative (weakening), and 38 positive (intensifying).

Using the TRMM data (spatial resolution $0.25^\circ$ latitude $\times$ $0.25^\circ$ longitude), for each $k$th position of the storm center in EBTRK, we established the dependence of precipitation on distance $r$ from the storm center, with $\overline{P}_{k}(r_i)$ defined as the mean precipitation in all grid cells with $r_i < 25 i$ km, $1 \le i \le 120$, $r_i \equiv 25 (i-1)+12.5$ km. Examples of precipitation distributions for individual hurricanes are given in Fig.~11 of \citet{ar17}.

The maximum value of thus obtained $\overline{P}_{k}(r_i)$ and the radius $r_i$ corresponding to this maximum were defined as $\overline{P}_{k}(r_P)$ and $r_P$, respectively, for the $k$th record. Additionally, $\overline{P}_{k}(r_i)$ for which $25 (i-1) \le r_m < 25 i$ km was defined as precipitation $\overline{P}_{k}(r_m)$ at the radius of maximum wind $r_m$ for storms where $r_m$ was known. To enable numerical comparison between precipitation and intensification rates\footnote{For example, one mm of water per hour (multiplied by $\rho_l g$, where $\rho_l$ is the density of liquid water) is equivalent to $2.4$ hPa per day.}, we expressed precipitation in hPa~day$^{-1}$.

To partially account for the different temporal resolutions of EBTRK (6 hours) and TRMM (3 hours), we compared $I_k$ with $P_m \equiv [\overline{P}_{k}(r_P)+\overline{P}_{k+1}(r_P)]/2$ and $P_r \equiv [\overline{P}_{k}(r_m)+\overline{P}_{k+1}(r_m)]/2$. However, the results were found to be largely insensitive to extending the time scale used to define the intensification rate to 12 hours, defining it as $I_k \equiv -2(p_{k+1}-p_{k-1})$, while comparing it with the 3-hour precipitation at each $k$th position of the storm. In all cases, precipitation was found to be similar in magnitude to the intensification rate.

\begin{figure*}[tb]
\begin{minipage}[p]{0.8\textwidth}
\centerline{\includegraphics[width=1\textwidth,angle=0,clip]{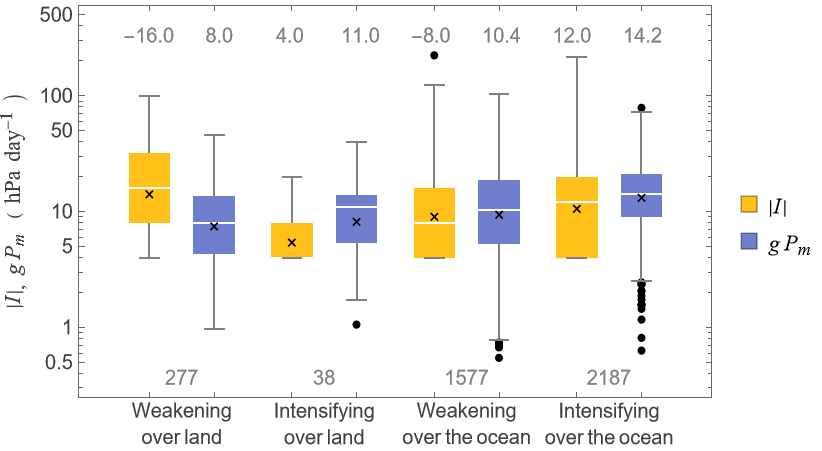}}
\end{minipage}
\caption{
Intensification rates taken by absolute value $|I|$ and maximum precipitation $gP_m$ in intensifying and weakening storms on land and over the ocean (shown for comparison). Numbers of storms in each group are shown along the lower horizontal axis. Medians of $I$ and $gP_m$ are shown along the upper horizontal axis. Note the logarithmic scale on the vertical axis. Crosses and dots show mean values and outliers, respectively.
}
\label{bwc}
\end{figure*}

\section{Results}

The median, lower and upper quartiles for  storms weakening over land are $I = -16$ ($-32$, $-8$) hPa~day$^{-1}$, while their maximum concurrent precipitation is $gP_m = 8$ ($4$, $14$) hPa~day$^{-1}$. For storms intensifying over land $I = 4$ ($4$, $8$) hPa~day$^{-1}$ and $gP_m = 11$ ($5$, $14$) hPa~day$^{-1}$ (Fig.~\ref{bwc}).

Intensification rates shown in Fig.~\ref{bwc} describe changes of the minimum surface pressure and thus correspond to the central pressure tendency in the axisymmetric model of \citet{sparks22a}. Minimum surface pressure by definition changes faster than the mean surface pressure.  Thus applying Eq.~\eqref{dpdt1} to the circle $r \le r_P$ and neglecting the precipitation term in Eq.~\eqref{dpdt1} should overestimate the absolute magnitude of $\pt \overline{p}/\pt t$ by at least  $gP_m/|I|\times 100\% \simeq 50\%$ in the weakening storms and by $275\%$ in the intensifying storms on land. For storms weakening and intensifying over the ocean, the corresponding inaccuracies would be remarkably similar at $130\%$ and $120\%$ (Fig.~\ref{bwc}). According to Eq.~\eqref{dpdt1a}, this implies that, in both weakening and intensifying storms, the vertically integrated air convergence is generally \textit{positive}, such that the sign of the pressure tendency is determined by the absolute magnitude of the precipitation rate.

\begin{figure*}[tb]
\begin{minipage}[p]{0.75\textwidth}
\centerline{\includegraphics[width=1\textwidth,angle=0,clip]{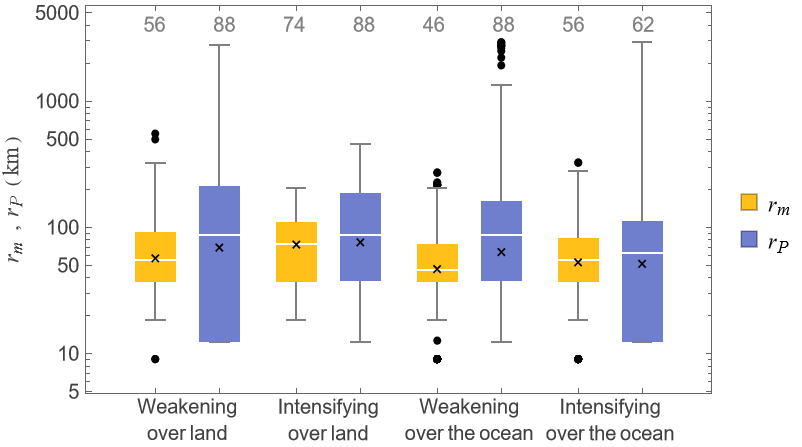}}
\end{minipage}
\caption{
Radius of maximum precipitation $r_P$ and radius of maximum wind $r_m$ in storms where $r_m$ is known. Numbers of
storms in each group are the same as in Fig.~\ref{bwc}. Medians of $r_P$ and $r_m$ are shown along the upper horizontal axis. Note the logarithmic scale on the vertical axis. Crosses and dots show mean values
and outliers, respectively.
}
\label{bwc2}
\end{figure*}

Figure~\ref{bwc2} shows that the radius of maximum precipitation, $r_P$, is somewhat larger than the radius of maximum wind, $r_m$, considered by \citet{sparks22a} in their model [Eq.~\eqref{st}]. By definition, precipitation within the radius of maximum wind, $\overline{P}_{k}(r_m)$, is lower than $\overline{P}_{k}(r_P)$. For storms over land, the median (lower, upper) quartiles of $gP_r$ were $4.9$ ($2.6$, $10$)~hPa~day$^{-1}$ for weakening storms and $6$ ($3.3$, $10$)~hPa~day$^{-1}$ for intensifying storms. The median (lower, upper) quartiles of $gP_r/|I|\times 100\%$ were $32$ ($13$, $69$)\% and $120$ ($36$, $220$)\% for weakening and intensifying storms, respectively.

Given such large inaccuracies, and under the reasonable assumption that the numerical model used by \citet{sparks22a} generated realistic precipitation, ignoring the mass sink to assess the pressure tendency from the net radial inflow alone could not produce meaningful results. Indeed, \citet{sparks22a} noted that {\textquotedblleft}the density-weighted column-integrated radial wind speed{\textquotedblright}  $\mathcal{U}_r$ {\textquotedblleft}was found to be unreliable when evaluated directly using instantaneous model output{\textquotedblright}, i.e., via its defining Eq.~\eqref{Ur}. Instead, to obtain reasonable agreement between their physical model and numerical simulations, \citet{sparks22a} had to resort to a circular approach -- first diagnosing $\chi$ from the mean pressure tendency using Eq.~\eqref{dpdt1} (thus implicitly accounting for precipitation) and then explaining  the pressure tendency using the thus-derived $\chi$ in their model.

In other words, {\textquotedblleft}column speeds{\textquotedblright} $\mathcal{U}_r$, $\chi$ and  $\chi_0$ shown in Fig. 4a,c, Fig. 6a,c and Table~1 of \citet{sparks22a} are not the actual mean column speeds but a surrogate variable $\mathcal{U}_s$ defined from the condition
\beq\label{Us}
\frac{\pt \overline{p}}{\pt t} \equiv -2p(r) \frac{\mathcal{U}_s(r)}{r}  ,
\eeq
such that
\beq\label{Us}
\mathcal{U}_s \equiv \mathcal{U}_r + \frac{r g \overline{P}}{2 p(r)}, 
\eeq
with the second term making a major contribution to the sum and of the opposite sign than the mean column speed in weakening storms. This calls for a re-consideration of the model{\textquoteright}s physical basis.

The mean column speed, $\chi$, as acknowledged by \citet{sparks22a}, is a non-observable quantity whose absolute magnitude is at least two orders of magnitude smaller than the actual radial velocities associated with inflow and outflow. In contrast, precipitation can be directly retrieved from observations. \citet{sparks22a} further suggested that their model could be extended to tropical cyclones over the ocean. For such cases, the similarity between $|I|$ and $gP_m$ is even more pronounced (see Fig.~\ref{bwc}), which led us to propose that precipitation may play a central role in both the intensification and de-intensification of storms \citep{snk24}. Because precipitation is expected to increase as the radius of maximum wind contracts and vertical velocities intensify, this mechanism could also explain the dependence of intensification and weakening rates on storm radius reported by \citet{sparks22b} using their model.

\section{Pressure adjustments related to precipitation}

\begin{figure*}[h!]
\begin{minipage}[p]{0.9\textwidth}
\centerline{\includegraphics[width=1\textwidth,angle=0,clip]{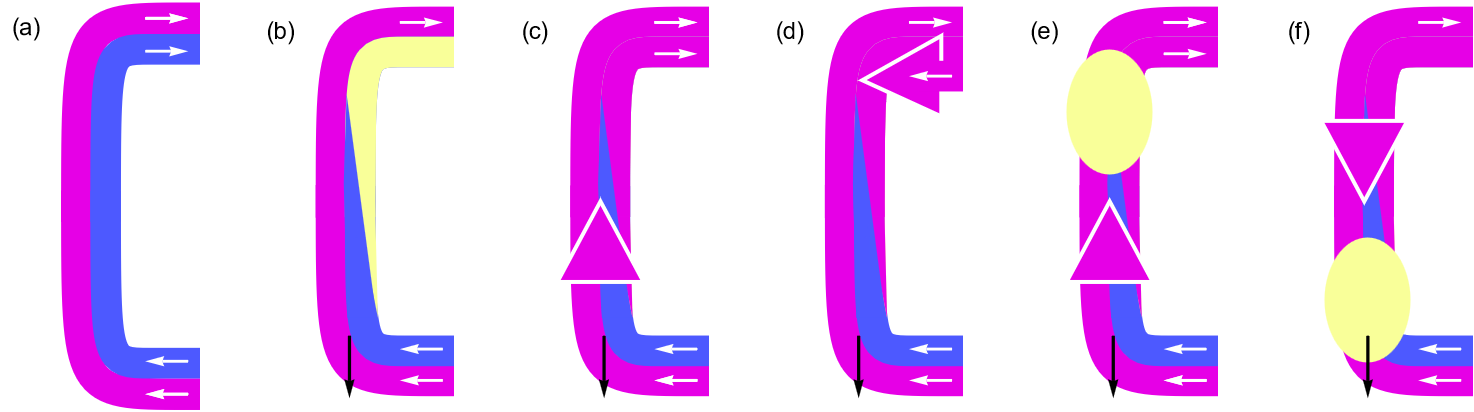}}
\end{minipage}
\caption{\normalsize
Thought experiments illustrating the role of pressure adjustments in shaping precipitation influence on storm intensification rate. Non-condensable and condensable gases are painted pink and blue, respectively. Thin white (black) arrows indicate inflow into, and outflow of gas (condensate) from, the column. Big triangles indicate the direction of pressure adjustment. Yellow spaces indicate pressure perturbations.
}
\label{fig3}
\end{figure*}

A thought experiment can help visualize how precipitation can impact intensification. Consider a steady-state circulation with a non-condensable tracer gas (Fig.~\ref{fig3}a). The radial inflow is equal to outflow and $\pt \overline{p}/\pt t = 0$. Now let us imagine that we begin to condense the tracer as it ascends, removing the condensate with precipitation (the black arrow in Fig.~\ref{fig3}b).  At the same time, we will {\it not} allow for any change in the flow velocity. In this imaginary case shown in Fig.~\ref{fig3}b, the flow remains steady: there is less tracer leaving the column as gas, but this reduction of the outflow is exactly compensated by precipitation. Precipitation {\it per se} does not lead to either intensification or de-intensification. It just escorts the condensed vapor from the column through another exit.

However, without any flow adjustment, we would have obtained a strongly non-hydrostatic column with uncompensated vertical pressure difference $\Delta p$ of the order of the partial pressure of water vapor $\Delta p \sim p_v \sim 30$~hPa. Had such $\Delta p$ persisted, we would have observed vertical velocities in excess of $50$~m~s$^{-1}$, which is clearly not the case in tropical storms. This means that precipitation {\it must} be accompanied by pressure adjustments. If this adjustment occurs in the vertical (Fig.~\ref{fig3}c), the pressure deficit in the upper atmosphere will be compensated, and the outflow restored up to (maximally) its unperturbed value. With the outflow and inflow again compensating each other, the surface pressure will fall at the (maximum) rate equal to precipitation\footnote{For example, Hurricane Milton 2024 that underwent rapid intensification at $84$ hPa~day$^{-1}$, should have had a precipitation maximum of at least $35$ mm~hour$^{-1}$. Reconnaissance flights into Milton recorded maximum local precipitation in excess of $30$  mm~hour$^{-1}$ and up to over $60$ mm~hour$^{-1}$, see \url{https://tropicalatlantic.com/recon/recon.cgi?basin=al\&year=2024\&product=hdob\&storm=Milton\&mission=16\&agency=AF\&ob=10-09-010230-38-910.3-140-164}. }.  If the pressure adjustment occurs in the horizontal (Fig.~\ref{fig3}d), this can lead to an additional reduction of the outflow. In this case the storm will de-intensify at a rate again (at maximum) determined  by precipitation.

Another way to look at the pressure adjustment problem is as follows. Is it possible to remove condensate from the atmosphere without disturbing the hydrostatic equilibrium? Consider again a hydrostatic column, where condensation and precipitation take place. At height $z$ we have
\beq\label{eq9}
- \frac{\pt p}{\pt z} = \rho g = \frac{p}{h}, \quad h\equiv \frac{RT}{Mg}, 
\eeq 
where $R$ is the universal molar gas constant, and the ideal gas equation of state $p = (\rho/M) RT$ is taken into account. For simplicity we ignore the difference in the molar masses $M$ of water vapor and dry air. Assuming air convergence to be negligible, the continuity equation reads
\beq\label{eq10}
\frac{\pt \rho}{\pt t} = \dot{\rho}.
\eeq 
Assuming also that temperature $T$ does not change, the rate of pressure fall is determined solely by the condensation rate, i.e., by the pressure sink $\dot{p}$ (W~m$^{-3}$): 
\beq\label{eq11}
\frac{\pt p}{\pt t} \simeq  \frac{RT}{M}\frac{\pt \rho}{\pt t}= gh \dot{\rho} \equiv \dot{p} < 0.
\eeq 
Differentiating Eq.~\eqref{eq9} with respect to time and taking  into account Eq.~\eqref{eq11}, we obtain
\beq\label{eq12}
 \frac{\pt \ln \dot{p}}{\pt z} =\frac{\pt \ln p}{\pt z} = - \frac{1}{h} .
\eeq 
This means that the removal of condensate from the atmospheric column will not disturb the hydrostatic equilibrium if the pressure sink is proportional to pressure itself, i.e., if  
\beq\label{k}
\dot{p}(z) = k p(z), 
\eeq
where $k$ is independent of altitude $z$. 

In the real atmosphere, condensation rate $\dot{\rho}$ is known to be determined by the material derivative of the water vapor mixing ratio \citep[e.g.,][their Eq.~(6)]{bryan09a}, to which the product of the vertical velocity and the vertical gradient of the water vapor mixing ratio makes a major contribution. Accordingly, the (steady-state) pressure sink $\dot{p}$ is approximately given by $\dot{p} = w p \pt \gamma/\pt z$ \citep[e.g.,][their Eq.~(18)]{jetp12}, such that $k = w \pt \gamma/\pt z$, where $\gamma \equiv p_v/p$ is the ratio of water vapor partial pressure $p_v$ to total pressure $p$. There are no grounds to expect the function $w \pt \gamma/\pt z$ to generally be a constant with respect to $z$.  While at higher temperatures $\pt \gamma/\pt z$ can be approximately constant below $z \lesssim 10$~km as governed by moist thermodynamics \citep[e.g.,][their Fig.~2]{jas13}, the vertical velocity $w$ can in principle change arbitrarily over $z$.

If $k(z)$ has a maximum and condensation is predominantly concentrated in the upper atmosphere (Fig.~\ref{fig3}e), the resulting pressure adjustment will be directed mainly upward, leading to storm intensification (convective bursts preceding rapid intensification may be an example of this regime). If, by contrast, $k(z)$ is maximized and condensation is more intense in the lower atmosphere, the pressure adjustment will proceed downward, suppressing upwelling and potentially resulting in storm de-intensification (Fig.~\ref{fig3}f). At some intermediate height of the condensation maximum, the intensification rate can become zero, corresponding to a steady storm state\footnote{With vertical velocity freely varying with height, these conclusions do not depend on our neglect of the difference in the molar masses of water vapor and dry air made when deriving Eq.~\eqref{eq12}. Whatever the exact form of Eq.~\eqref{eq12}, it does not contain vertical velocity.}.

In view of $|I| \simeq gP$, the concepts illustrated in Fig.~\ref{fig3} are therefore directly relevant to storm intensification, yet they remain largely unexplored.

Figure~\ref{fig3} illustrates that, although the pressure tendency is determined by the precipitation mass sink, it pertains primarily to dry air, since the void created by water vapor condensation is filled by atmospheric air in which dry air dominates. This can also be seen by considering the mass budgets of atmospheric water and dry air separately, as suggested by a reviewer. Using the mass budget equation for water in Eq.~\eqref{spt},
\beq\label{W}
\frac{\pt M_v}{\pt t} + \int_{0}^{H} \mathrm{div} (\rho_v \mathbf{u}) dz = E - P ,
\eeq
where $M_v$ is the total mass of precipitable water in the atmospheric column, we obtain
\beq\label{spt2}
\frac{\pt p}{\pt t} - g \frac{\pt M_v}{\pt t} = -g\int_{0}^{H} \mathrm{div} (\rho_d \mathbf{u}) dz .
\eeq
Since the surface pressure of an intensifying tropical cyclone typically decreases by several tens of millibars \citep{chavas17}, whereas the total precipitable water content increases by only about 20~mm \citep[][their Fig.~4a]{ar17}---corresponding to a surface pressure increase of roughly 2~mb---the second term on the left-hand side of Eq.~\eqref{spt2} can, in most circumstances, be neglected. This implies that the surface pressure tendency in tropical storms is, to high precision, determined by changes in the dry-air mass\footnote{The mean column velocity may then be defined by Eq.~\eqref{Ur}, with $\rho=\rho_d + \rho_v$ replaced by the dry-air density $\rho_d$. Under this reformulation, the model of \citet{sparks22a} would be formally valid. However, although $\rho$ and $\rho_d$ differ by only about 1\%, the column-mean velocities obtained using $\rho_d$ and those obtained using $\rho$ with the precipitation term neglected can differ by more than 100\% and may even have opposite signs. It is therefore essential to specify the definition of air density in Eq.~\eqref{Ur} precisely in order to ensure the internal consistency of the model. Because the mean column velocity is defined using $\rho$ in both \citet{sparks22a} and the companion paper by \citet{sparks22b}, this requirement is not satisfied.}.

Taken together, the data in Fig.~\ref{bwc} and Eq.~\eqref{spt2} indicate that, in intensifying storms, the tendency of dry-air mass in the atmospheric column is of the same order as the precipitation rate within the radius of maximum precipitation. In other words, within the core precipitation region of the storm, the dry-air content decreases at a rate comparable to that at which precipitation removes water from the column\footnote{The central pressure tendency overestimates the mean pressure tendency; consequently, the relative contribution of precipitation to the mean pressure tendency is greater than indicated by the data in Fig.~\ref{bwc}. Additionally,
the coarse spatial resolution of TRMM 3B42 is expected to underestimate precipitation in the most intense storms with small radii of maximum wind.}.

We argue that this correspondence warrants further investigation. By removing water vapor from the column, precipitation perturbs the vertical pressure distribution and induces compensating air motions that would be absent in a dry atmosphere. Evidence for the dynamical importance of this precipitation-induced mass sink is provided by both observations and numerical models. Pressure gradients that drive atmospheric motion act equally on dry air and water vapor. The presence of positive vertically integrated air convergence in cyclones therefore implies that upper-level pressure gradients, together with centrifugal forces, are unable to export air from the storm core at a rate sufficient to offset the low-level moisture inflow. As a result, relatively small changes in precipitation rate can shift a cyclone from intensification to weakening. Consistent with this interpretation, \citet{su2020} found that, across all intensity categories from tropical depressions to major hurricanes, intensifying storms exhibit a nearly universal precipitation surplus of approximately 2~mm~h$^{-1}$ in the storm core relative to weakening storms. Further support comes from modeling studies showing that suppressing precipitation fallout, while retaining latent heat release, strongly weakens storm dynamics \citep{bryan09a,wang21}. In a modeling study, \citet{smith2021} found that the development of a low-level outflow is associated with storm weakening. This behavior can also be interpreted in terms of the role of precipitation: unlike a high-level outflow, a low-level outflow is not conducive to substantial precipitation. As upward motion weakens, precipitation decreases, and the storm correspondingly weakens.

At present, these patterns lack a quantitative theoretical explanation. We suggest that focusing on surface pressure tendency provides a promising framework for understanding tropical cyclone dynamics, provided that the precipitation-induced mass sink is incorporated fully and consistently.

\section*{\large Acknowledgments}
{\large We thank three anonymous reviewers and Editor Dr. Daniel Stern for their comments and suggestions. Work of A.M. Makarieva is partially funded by the Federal Ministry of Education and Research (BMBF) and the Free State of Bavaria under the Excellence Strategy of the Federal Government and the L\"ander, as well as by the Technical University of Munich -- Institute for Advanced Study.}

\section*{\large Data statement}
The raw data utilised in this study were derived from the following resources available in the public domain:
\url{https://disc.gsfc.nasa.gov/datasets/TRMM\_3B42\_7/summary} and \url{https://rammb2.cira.colostate.edu/research/tropical-cyclones/tc\_extended\_best\_track\_dataset/}. Precipitation data for all analyzed storms are available at \url{https://zenodo.org/records/10577109}.

\appendix
\section{Response to Reviewers and Editor}

{\color{black} \it
\noindent

\noindent
\textbf{Manuscript Number:} JAS-D-24-0178

\noindent
\textbf{Authors:} Makarieva, A. M. and A. V. Nefiodov

\noindent
\textbf{Title:} Comments on “A physical model of tropical cyclone central pressure filling at landfall” by
Sparks and Toumi

\noindent
\textbf{Decision:} Reject

\noindent
\textbf{Summary}

\noindent
The authors present an overview of the surface pressure tendency in a hydrostatic
atmosphere and contributions to the surface pressure tendency from mass convergence and
precipitation. Their study primarily critiques the absence of a precipitation mass sink in a
simplified model of surface pressure filling in landfalling tropical cyclones introduced by Sparks
and Toumi (2022). The ramifications of neglecting contributions from the precipitation mass
sink on the surface pressure tendency are summarized, and thus the authors provide a valid
critique of the Sparks and Toumi model. However, I cannot recommend the manuscript for
publication with the Journal of the Atmospheric Sciences in its current state.

{\color{blue}\normalfont
We are grateful to the reviewer for the helpful comments and for recognizing the merit of our critique.}

\vspace{0.3cm}\noindent
The manuscript abruptly begins with a textbook introduction to the surface pressure
tendency in a hydrostatic atmosphere within the context of a cylinder encompassing a tropical
cyclone. Substantial context from previous literature is missing from the Introduction section that
would enable the study to stand on its own in the absence of a response from Sparks and Toumi.

{\color{blue}\normalfont
In the revised manuscript, we do not derive the pressure tendency equation,  but refer to the previous literature.
We have also incorporated specific examples from the published literature, which were not referenced by Sparks and Toumi, that demonstrate how neglecting the precipitation term in the pressure tendency equation can lead to substantial errors. For example, in some cases this omission can result in diagnosing a storm as weakening when it is actually rapidly intensifying.
}

\vspace{0.3cm}\noindent
Beginning on Line 119, the discussion surrounding thought experiments about how the
precipitation mass sink might impact tropical cyclone intensification is superfluous to the
critique of the Sparks and Toumi study. The authors do not map the conclusions of their thought
experiments to any specific elements of the Sparks and Toumi model. In general, much of the
discussion beyond Line 119 is explained in greater detail by the authors’ self-referenced study
Makarieva and Nefiodov (2024).

{\color{blue}\normalfont
We agree that it is not directly related to the critique of the Sparks and Toumi study and have therefore removed this part from the revised manuscript, 
though we did so with some regret, as we believe it offered a broader and more interesting perspective on the main point concerning the 
straightforward omission of a major term from the mass balance.
}

\vspace{0.3cm}\noindent
In general, it is worth noting that the thought experiment and analyses presented herein
are further limited by the prerequisite assumption of a hydrostatic atmosphere to relate the
precipitation mass sink to the surface pressure tendency. The tropical cyclone eyewall can
substantially deviate from hydrostatic balance, and thus the pressure of the fluid is subject to
linear and nonlinear perturbations associated with velocity gradients that are not considered in
the framework proposed by the authors. It is unclear how nonhydrostatic processes in the tropical
cyclone eyewall might complicate the proposed causal relationship between the precipitation
mass sink and the surface pressure tendency in a hydrostatic atmosphere. I encourage the authors
to consider the implications of nonhydrostatic processes on their framework in future studies.

{\color{blue}\normalfont
Thank you for this comment on our thought experiment and the accompanying analysis. While we have excluded this part from the revised manuscript, we believe it presents an original insight: 
namely, that independent of the exact nature of the vertical pressure distribution,
how/whether condensation will disturb this distribution, initiating pressure adjustments, will depend on the vertical profile of vertical velocity.
The non-hydrostaticity of the eyewall does not alter this conclusion—it would only affect the magnitude of $h$ in Eq.~(12) of our original manuscript (available at \url{https://arxiv.org/abs/2410.14717v1}).
}

\vspace{0.5cm}
\noindent
Apr 07, 2025

\noindent
Ref.: JAS-D-24-0178

\vspace{0.3cm}\noindent
Dear Dr. Makarieva,

\vspace{0.3cm}\noindent
First, I apologize for the length of time from your submission to a decision, thank you for your patience.  I’m now in receipt of a review of your Comment on “A Physical Model of Tropical Cyclone Central Pressure Filling at Landfall.” Although the reviewer sees merit in your critique of the Sparks and Toumi model, they recommend rejection of the Comment, on the grounds that it is unable to stand on its own as a manuscript in the absence of a Reply, which the authors of the original paper have declined to provide.  The reviewer argues that substantial context from previous literature is missing from the introduction, which would be necessary for a reader to consider this as an original manuscript.  And they further note that the proposed thought experiment and related discussion are superfluous to the critique of the Sparks and Toumi study.

\vspace{0.3cm}\noindent
I have also carefully read through both your Comment and the original paper that you are commenting on, and while I agree that precipitation (which is neglected in the framework of Sparks and Toumi) may have some effect on the surface pressure tendency, I believe that this Comment is unable to demonstrate that (1) this is a leading order effect and (2) that the neglect of this term in Sparks and Toumi relates to their inability to directly evaluate the column-mean inflow speed.  For (1), this is primarily because the time tendency of vertically integrated water (in all its forms) is not in general given simply by the sum of local evaporation and precipitation.  The tendency is also influenced by horizontal advection, which cannot be neglected if you are trying to make an argument about the effects of water on the total hydrostatic pressure.  For a steady TC over the ocean, for example, the mass of water being lost to precipitation is continuously being replenished by the net
effect of local surface fluxes and inward advection, and so the local surface pressure is not changed by this “loss” of water.  The sum of the water budget may change during landfall and so it is possible that changes in the precipitation rate may influence the rate of pressure change as you suggest, but the analyses in the Comment are not able to show this on their own.  For (2), although it is possible that the neglect of water impedes an accurate determination of the column-mean inflow in a numerical model, I think it is more likely that the errors are instead dominated by the difficulties in calculating a term that is the small residual of components that are several orders of magnitude larger.  Further, Sparks and Toumi also evaluated simulations with all moisture removed, and it is implied that the same problem occurred, which would argue against the supposition that this issue is driven by the neglect of moisture in the theoretical framework.

\vspace{0.3cm}\noindent
Based on both the recommendation of the reviewer and my own evaluation, I unfortunately must reject this Comment manuscript for publication.  I note that there is overlap between this Comment and your manuscript “Condensation Mass Sink and Intensification of Tropical Storms”, previously considered for JAS.  If you are able to incorporate your critique of Sparks and Toumi into that manuscript while also satisfactorily addressing the issues (which overlap with what I discuss above) raised by the reviewers and myself, then I would consider a resubmission.

\vspace{0.3cm}\noindent
If you choose to address the problems identified by the reviewers and resubmit your manuscript, please upload it to the Editorial Manager as a new submission and follow the AMS Resubmission Requirements at: www.ametsoc.org/PubsResubmission. You will need to declare it as a resubmission and include a Response to Reviewers.

\vspace{0.3cm}\noindent
Along with your substantially revised manuscript, please upload a point-by-point response that satisfactorily addresses the concerns and suggestions of each reviewer and the Editor. To help us to assess your revisions, our journal recommends that you cut-and-paste the reviewer and Editor comments into a new document. As you would conduct a dialog with someone else, insert your responses in a different font, font style, or color after each comment. If you have made a change to the manuscript, please indicate where in the manuscript the change has been made. (Indicating the line number where the change has been made would be one way, but is not the only way.) Although it is not required by our journal, you may wish to include a tracked-changes version of your manuscript. You will be able to upload this as "additional material for reviewer reference." Should you disagree with any of the proposed revisions, you will have the opportunity to explain your rationale in your
response. No separate cover letter to me is needed unless it contains essential information that does not appear in your reply.

\vspace{0.3cm}\noindent
Thank you for your interest in Journal of the Atmospheric Sciences. Please contact me at Stern.JAS@ametsoc.org if you have any questions. We wish you a more positive result with future endeavors.

\vspace{0.3cm}\noindent
Best regards,

\vspace{0.3cm}\noindent
Daniel Stern

\noindent
Editor

\noindent
Journal of the Atmospheric Sciences

\vspace{0.5cm}

{\color{blue}\normalfont
\noindent
26 May 2025
\vspace{0.5cm}

\noindent
Dear Dr. Stern,
\vspace{0.5cm}

\noindent
Thank you for your consideration of our previous submission, titled {\textquotedblleft}Comments on {\textquotedblleft}A Physical Model of Tropical Cyclone Central Pressure Filling at Landfall{\textquotedblright} by Sparks and Toumi.{\textquotedblright} We appreciate the feedback from the reviewer and editor. After reflecting on the comments received, we respectfully request that you consider this revised version of our commentary. We believe there are compelling reasons to revisit the decision, which we outline below.

\itemize{

\item
    Sparks and Toumi based their model for the surface pressure tendency on the continuity equation for moist air, but neglected the source term---evaporation minus precipitation. This introduces a significant error, as this omitted term has been shown in previous literature on modeled cyclones, not cited by Sparks and Toumi, to be of the same order of magnitude as the surface pressure tendency itself. 

\item
    In our commentary, we present original analyses of observational data demonstrating that the neglected source term is indeed of the same order of magnitude as the retained term not just in models but in real-world cyclones. This empirical evidence highlights a substantial limitation in the Sparks and Toumi{\textquoteright}s model and renders it grossly inaccurate.
We show, using published data not
referenced by Sparks and Toumi, that in some cases neglecting the precipitation term will result in diagnosing the storm as weakening while it is in fact rapidly intensifying.

\item
    The argument that the water vapor budget includes moisture convergence in addition to the source term is, in itself, correct. However, it is not relevant to Sparks and Toumi{\textquoteright}s expression for pressure tendency, which is based on the mass budget of moist air as a whole. Our critique pertains to that same total mass budget, not to the water vapor budget alone. That such a misunderstanding arose, however, suggests that other readers may also be confused, reinforcing the value of clarifying this issue explicitly.

\item
    The suggestion that Sparks and Toumi could not directly diagnose the primary parameter of their model, the column-mean radial speed, for reasons other than the omission of a critical term remains speculative, especially in the absence of a response from the authors. The fact that they used the same procedure for their dry simulation does not necessarily indicate that they encountered the same difficulty in the dry case, but may simply reflect their decision to apply a uniform methodology across simulations.

\item
    The authors declined to respond to a critique that an independent reviewer acknowledged as valid. We are concerned that this absence of engagement could be misinterpreted as a reason to disregard a scientifically important 
issue\footnote{In a previous instance, we submitted a critical comment to another journal where the authors, while initially willing, ultimately also declined to provide a response. However, in that case, they explicitly acknowledged the validity of our critique in their communication with the editor, and the journal published our comment without a reply (\url{https://doi.org/10.1175/JCLI-D-14-00592.1}). We believe this reflects a transparent and constructive editorial practice. While authors may choose not to respond, silence in the face of a valid, peer-recognized critique should not be a reason to dismiss such commentary.}.
}

}

{\color{blue}\normalfont
We acknowledge your observation regarding some overlap between this commentary and our earlier submission {\textquotedblleft}Precipitation mass sink and intensification of tropical storms{\textquotedblright}. While that submission was ultimately rejected, it received favorable comments from three reviewers recommending major revisions. Although it briefly mentions Sparks and Toumi, the earlier manuscript includes extensive independent content. Should we be in a position to revise and resubmit it, it would be judged on its own merit. However, if rejected again, the critique of Sparks and Toumi{\textquoteright}s paper---central to the present commentary---would remain unpublished and unaddressed.

Therefore, we believe it is valuable to treat this commentary as a standalone submission. It is concise, presents original data, directly engages with Sparks and Toumi{\textquoteright}s publication, and has been revised to address all reviewer concerns, including the addition of contextual background and the removal of the content judged by the reviewer as extraneous.

Furthermore, we note that the editorial policies of the American Meteorological Society explicitly encourage the correction of errors in the published record when they are identified. We believe that publishing this comment, which addresses a substantive methodological issue, aligns with the principles of good science.

We thank you for your time and consideration, and we hope you find the revised manuscript appropriate for publication as a commentary in Journal of the Atmospheric Sciences.

Sincerely,

Anastassia Makarieva
}

\vspace{0.5cm}\noindent
Review report of JAS-D-25-0078: "Comments on "A Physical Model of Tropical Cyclone
Central Pressure Filling at Landfall" by Sparks and Toumi"

\vspace{0.3cm}
\noindent
Summary: This paper presented a reasoning with observational support that the mass sink due
to precipitation is important for a tropical cyclone central pressure tendency.
\vspace{0.1cm}

\noindent
Evaluation: while I do not think the authors' reasoning is mathematically incorrect, I do think it can be physically misleading, as I will explain more below.
\vspace{0.1cm}

\noindent
Major comment:
\vspace{0.1cm}

\noindent
I think it would be physically more intuitive to write your Eq. (6) as
\vspace{0.1cm}

\hspace{-0.8cm}
\includegraphics[width=150mm]{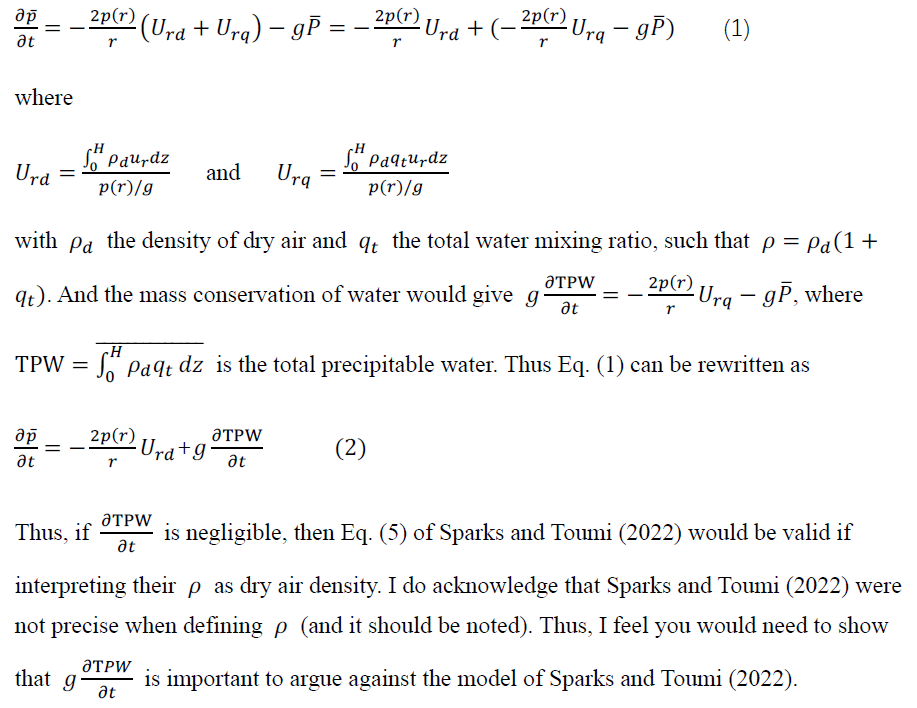}
\vspace{0.1cm}

{\color{blue}\normalfont
We thank the reviewer for their time and insightful comments. We fully agree that, if $\rho$ were replaced by the dry-air density $\rho_d$, the model of Sparks and Toumi would be formally valid, and we have incorporated this reasoning into the revised text of the Comment (the last but one page).

However, as also noted by the reviewer, $\rho_d$ is not mentioned anywhere in the governing equations of the model. Although total air density and dry-air density differ by only about one percent, the key diagnostic quantity in the model—the column-mean speed—can differ by hundreds of percent and may even change sign depending on which density is used. This therefore goes beyond a minor or cosmetic issue.

Moreover, while the model could in principle be reinterpreted by replacing $\rho$ with $\rho_d$, the numerical analyses presented in the paper cannot be modified retroactively. In particular, if the difficulty in diagnosing the mean column speed from the numerical model is related to its definition in terms of $\rho$, this connection should be stated explicitly.

In light of these considerations, we suggested to the Editors that publication of our Comment would help clarify the issue, and/or that the authors be invited to consider a corrigendum.
}

\vspace{0.3cm}\noindent
In your present form of comment, you are suggesting that the pressure change is contributed
substantially by precipitation. This is against my intuition (obtained from previous works) that
the tropical cyclone central pressure drop is due to adiabatic warming of subsidence air (so
that the eye is clear). 

\vspace{0.3cm}
{\color{blue}\normalfont
We agree that subsidence and associated adiabatic warming play an important role in shaping the thermodynamic structure of the eye. However, subsidence of warm air does not, by itself, imply a decrease in surface pressure; warm air can also descend in regions of high pressure, such as anticyclones during heatwaves. The reviewer’s intuition is therefore best interpreted in terms of a constant isobaric height: if an isobaric surface connects regions of different temperature, the warmer region will have lower surface pressure. In tropical cyclones this concept is often associated with the height of the tropopause. In practice, however, surface pressure is only weakly constrained by temperature alone, owing to its strong sensitivity to the vertical temperature profile (see Fig.~3 of Makarieva et al., 2015, Journal of Climate).

As a result, the pressure within the eye is also dynamically controlled, with the eye approximately following solid-body rotation set by the eyewall wind speed. Because the circulation in the eye involves negative work (subsidence of positively buoyant warm air), the eye is dynamically controlled by the eyewall, where positive work is generated (and not vice versa).

Irrespective of the mechanism responsible for the pressure reduction in the eye, its contribution represents only a small fraction of the total pressure drop in a tropical cyclone. The eye radius is typically about half the radius of maximum wind, while the radius of maximum precipitation is larger still. Consequently, the pressure within the eye makes only a minor contribution to the mean surface pressure over the storm core. The central pressure tendency therefore serves primarily as a proxy for the mean pressure tendency, to which it can be related, for example, via the Holland pressure profile. It is this mean pressure tendency that is directly affected by precipitation through mass removal.

}

\vspace{0.3cm}\noindent
In fact, I would argue that the moisture mass is brought in by the inflow 
but then the condensed hydrometeors fallout, so the net effect on surface pressure would be
negligible.

\vspace{0.3cm}{\color{blue}\normalfont
We agree that most of the imported water vapor subsequently condenses and precipitates, such that the total pressure tendency is very accurately described by the dry-air pressure tendency. However, this does not imply that the resulting surface pressure tendency is not dynamically influenced by the precipitation mass sink.

In our initial submission, we included a conceptual figure illustrating how hydrostatic pressure adjustment associated with water vapor removal aloft can induce additional upward motion and a subsequent outflow of predominantly dry air, in amounts related to the removed water vapor (see Fig.~3 in the arXiv version \url{https://arxiv.org/pdf/2410.14717}). This mechanism offers a possible physical interpretation of the remarkable observation that the dry-air pressure tendency is comparable in magnitude to the precipitation rate.

One reviewer considered this material peripheral to the specific critique of Sparks and Toumi, and we therefore agreed to remove it.

In the revised Comment, we instead present a concluding argument—based on observational evidence and previously published literature—that provides additional support for our proposition that the dynamic influence of the precipitation mass sink on the mean surface pressure tendency may be significant and merits further investigation.
}

\vspace{0.5cm}

\noindent
\textbf{Review of Manuscript JAS-D-25-0078}

\vspace{0.3cm}\noindent
\textbf{Summary}

\vspace{0.3cm}\noindent
This manuscript presents a "Comment" on the paper "A Physical Model of Tropical Cyclone
Central Pressure Filling at Landfall" by Sparks and Toumi (2022) (S\&T). The authors argue that
the S\&T model is fundamentally flawed due to its neglect of the mass sink from precipitation
when formulating the surface pressure tendency. The Comment uses theory, prior literature, and
an analysis of observational data to assert that precipitation is a first-order term in the pressure
tendency equation and that S\&T employed a circular methodology as a result of this omission.

\vspace{0.3cm}\noindent
\textbf{General Comments}

\vspace{0.3cm}\noindent
The authors of this Comment raise an important and potentially valid critique regarding the
physical basis of the S\&T (2022) model. Their argument—that the precipitation mass sink is a
critical component of the surface pressure tendency in tropical cyclones—is at least intuitive and
may be correct, though the role of mass removal due to precipitation remains largely unexplored
in the broader community (this does not necessarily diminish its potential importance).

\vspace{0.3cm}
{\color{blue}\normalfont
We thank the reviewer for their time and careful reading of the manuscript, and for noting that the Comment raises an important and potentially valid critique of the physical basis of the Sparks and Toumi (2022) model. We also appreciate the observation that the role of the precipitation-related mass sink in the surface pressure tendency has been relatively little explored in the broader literature.

The summary provided by the reviewer accurately reflects the scope and intent of our Comment. The manuscript does not introduce any new theory. Rather, our goal is to apply established theoretical principles (like mass conservation), prior literature, and observational evidence to assess whether the omission of the precipitation mass sink materially affects the surface pressure tendency formulation and the conclusions of S\&T. With this in mind, we note that the reviewer’s subsequent comments appear to focus primarily on concerns about a “novel theory,” rather than engaging directly with this central question.
}

\vspace{0.3cm}\noindent
Nonetheless, the manuscript’s own supporting arguments and methodology exhibit several
significant issues that undermine confidence in its conclusions. The proposed framework
appears to overlook key physical processes and spatial structures within hurricanes, and the
appropriateness of the selected observational datasets is questionable.

\vspace{0.3cm}\noindent
Given these limitations, I suggest that the authors consider developing this Comment into a full
article, ideally grounded in idealized numerical simulations to rigorously test their hypothesis. In
its current form, I recommend rejection, primarily due to deficiencies in the arguments, and
secondarily because the concept represents a novel “theory” that would be better suited to a
standalone article.

\vspace{0.3cm}
{\color{blue}\normalfont
We respectfully note that the characterization of the Comment as introducing a “novel theory” is difficult to reconcile with the reviewer’s own summary, which correctly describes the manuscript as a critique. Furthermore, as we clarify below, the concerns raised appear to pertain primarily to the assumptions and simplifications adopted in Sparks and Toumi (2022), which are the subject of our critique and therefore reinforce, rather than weaken, its motivation.
}

\vspace{0.3cm}\noindent
\textbf{Major Comments}

\vspace{0.3cm}\noindent
1. \textbf{The "Eye Conundrum" and the Role of Mass Convergence:} 
 The Comment's central
thesis is that precipitation is the dominant term in the pressure tendency equation.
However, this argument fails when considering the storm's eye. The eye is a region of
subsidence with little to no precipitation, yet it is where the lowest surface pressure and
the most significant pressure changes are observed. If one were to average the pressure
tendency equation over the eye alone, the precipitation term (gP) would be zero. In this
case, the pressure tendency must be driven entirely by the vertically integrated mass
convergence term, the very term that S\&T focused on. This suggests that the
Comment's conclusion is not general and that the relative importance of mass
convergence versus precipitation is highly dependent on the spatial area being
considered. The authors must address how their framework accounts for the observed
pressure changes in the largely precipitation-free eye.

\vspace{0.3cm}
{\color{blue}\normalfont
We respectfully note that the reviewer attributes to the Comment a claim that is not made therein, namely that the precipitation term is universally dominant in the surface pressure tendency equation. No such general statement is advanced in the manuscript.

Following the formulation of Sparks and Toumi (2022), our analysis is explicitly concerned with the region for which their model is constructed, namely the radius of maximum wind. It is in this region that we show the precipitation-related mass sink to be a leading-order contribution to the pressure tendency. The behavior of the pressure tendency within the eye, where subsidence dominates and precipitation is weak or absent, lies outside the scope of the S\&T framework being evaluated and therefore outside the scope of our critique.

That said, we fully agree with the reviewer that, if a novel theory were being proposed, it would indeed be essential to address the pressure tendency within the eye. We respectfully refer the reviewer to our response above to a related comment raised by another reviewer, where this issue is discussed.
}

\vspace{0.3cm}\noindent
2. \textbf{Oversimplified Physics:} The analysis is based on a simplified, hydrostatic framework
that neglects several potentially important physical effects in a heavily precipitating
system:

\noindent
-- \textbf{Hydrometeor Mass:} The mass continuity equation as presented accounts for
the mass of air and water vapor, but as far as I understand does not explicitly
account for the substantial mass of suspended and falling liquid/ice hydrometeors
within the atmospheric column. This problem should be addressed.

\noindent
-- \textbf{Precipitation Drag and Non-Hydrostatic Effects:} Intense precipitation exerts a
downward drag on the air and is associated with strong vertical accelerations.
These processes introduce significant non-hydrostatic pressure effects that are
not considered in the Comment's hydrostatic framework. These unaddressed
complexities weaken the claim that the presented equation is sufficient to fully
describe the pressure tendency.

\vspace{0.3cm}
{\color{blue}\normalfont
Our analysis adopts the same hydrostatic framework as Sparks and Toumi (2022), who relate surface pressure to the total mass of the atmospheric column. In this formulation, the column mass includes all constituents contributing to weight, including suspended and falling hydrometeors. Their presence therefore does not constitute an omission from the mass budget relevant to surface pressure. The inference that hydrometeors are omitted from our formulation is therefore not supported.

With respect to precipitation drag and non-hydrostatic effects, we agree that strong local vertical accelerations can occur in heavily precipitating regions. However, these effects represent small departures from hydrostatic balance when considering the vertically integrated column mass. Hydrometeors typically fall close to their terminal velocity, such that the drag they exert approximately balances their weight, and non-hydrostatic contributions do not materially affect the surface pressure diagnosed from column-integrated mass.

If these effects were large enough to invalidate the hydrostatic approximation, they would undermine not only the formulation of Sparks and Toumi (2022), but also the previous analyses of storm pressure tendencies cited in our Comment, all of which rely on hydrostatic equilibrium to relate surface pressure to column mass.
}

\vspace{0.3cm}\noindent
4. \textbf{Suitability and Limitations of the Observational Datasets:} The observational analysis
is undermined by the potential limitations of the chosen datasets.

\noindent
-- \textbf{TRMM Data:} The manuscript relies on the TRMM 3B42 product, which has a
relatively coarse spatial resolution (0.25° × 0.25°) for estimating rainfall within a
tropical cyclone’s core. The authors should justify the suitability of this dataset
and discuss the uncertainties it introduces, particularly in comparison with
higher-resolution products such as the TRMM Precipitation Radar (PR).

\noindent
-- \textbf{Best Track Data:} The 6-hourly temporal resolution of the best track dataset is
coarse for analyzing pressure tendency. Tropical cyclones can intensify or
weaken rapidly on much shorter timescales. Calculating a 12-hour pressure
change $(p_{k+1} - p_{k-1})$ may miss the periods of most rapid change and lead
to a temporal mismatch with the 3-hourly precipitation data, potentially
introducing a severe bias.

\vspace{0.3cm}
{\color{blue}\normalfont
The primary purpose of the observational analysis is to demonstrate that the precipitation term is non-negligible. The results obtained are fully consistent with previously published, peer-reviewed studies cited in the manuscript but not considered by Sparks and Toumi (2022). In this context, our findings reinforce existing evidence that precipitation contributes materially to the mean surface pressure tendency within the storm core. Accordingly, the justification for neglecting this term rests with Sparks and Toumi.

No observational dataset is without limitations. The TRMM 3B42 product has been extensively evaluated in tropical cyclone environments and shown to provide reliable estimates of area-averaged rainfall when compared with TRMM PR and rain-gauge observations. Our analysis focuses on spatially averaged precipitation over scales relevant to the column-integrated mass budget, rather than on resolving fine-scale rainfall structure within the storm core. 

We note separately that the coarse spatial resolution of TRMM 3B42 is expected to underestimate precipitation in the most intense storms with small radii of maximum wind. This implies
that the impact of precipitation in those storms is even greater than shown in our Fig.~1.

We respectfully disagree that the 6-hourly temporal resolution is too coarse for analyzing pressure tendency; for example, rapid intensification is commonly defined using pressure changes over one day or longer
(see, e.g., the study of Su et al. 2020 cited in the revised Commentary). Nevertheless, we agree that the time scales of precipitation and pressure tendency should be matched as closely as possible. In the revised Comment, we therefore reanalyzed the data using 6-hour intensification rates and the mean of two 3-hour precipitation estimates at successive storm positions. The revised results show no significant changes to the conclusions.
}

\vspace{0.3cm}\noindent
4. \textbf{Reliance on Non-Peer-Reviewed Literature:} A key part of the authors' concluding
argument, which posits a central role for precipitation in both intensification and
weakening, relies on a citation to their own 2024 study that is currently a preprint on the
arXiv server. Relying on non-peer-reviewed work, particularly one's own, to support a
primary conclusion is not good practice. The arguments in this Comment should stand
on their own merit, based on the evidence presented herein and on established,
peer-reviewed literature.

\vspace{0.3cm}
{\color{blue}\normalfont
We respectfully note that the arXiv preprint cited by the reviewer appears only in the penultimate sentence of the Comment and represents the sole reference to any broader or forward-looking theoretical context. It is not used to support the central arguments of the manuscript. The critique of Sparks and Toumi (2022) stands independently of this reference, and the conclusions regarding the role of precipitation in the surface pressure tendency do not rely on it in any way.

Citing arXiv preprints is legitimate in many fields of science, as preprints play an important role in promptly reporting ongoing work in the field. Specifically, in our case, were this single citation removed, the manuscript would contain no reference to any novel theory, and the assessment would necessarily center on the validity of the critique of S\&T. While we appreciate the reviewer’s interest in the broader implications briefly mentioned, we emphasize that the primary focus of the Comment is the unjustified omission of the precipitation term in the model of Sparks and Toumi (2022).
}

\vspace{0.5cm}

\noindent
Oct 24, 2025

\noindent
Ref.: JAS-D-25-0078

\vspace{0.3cm}\noindent
Dear Dr. Makarieva,

\vspace{0.3cm}\noindent
I am now in receipt of two reviews of your resubmitted Comment on “A Physical Model of Tropical Cyclone Central Pressure Filling at Landfall.”  Note that neither of these reviewers reviewed the original Comment.

\vspace{0.3cm}\noindent
Although Reviewer 1 sees some potential merit in your argument, they ultimately recommend rejection, primarily because of “several significant issues” that “undermines confidence” in your conclusions.  In my view, the most substantial of these issues is that as Reviewer 1 points out, your argument that the removal of mass by precipitation plays a dominant role in the pressure tendency within a TC simply doesn’t work when considering the eye, where precipitation is absent.

\vspace{0.3cm}\noindent
Reviewer 2 recommends Major Revision, but they raise a critical issue, which is that precipitation fallout is only physically meaningful as a mass sink if the moisture in the column is decreasing (non-negligibly) in time as a result.  The net change in water mass is likely to be the small residual of fallout, local evaporation, and moisture convergence, and during TC intensification, the total precipitable water (TPW) is likely increasing in time, not decreasing.  As Reviewer 2 points out, if the change in TPW is small, then Eq. (5) of Sparks and Toumi (2022) remains valid when rho is defined as dry air density.  Combined with the fact that Spark and Toumi don’t make a direct analysis of their Eq. (5), it isn’t really possible to say that their framework has necessarily neglected a leading order term.  In order to really show that their framework is invalid (or that the precipitation mass sink is a physically important driver of intensity change), idealized simulations are
necessary, as suggested by Reviewer 1.

\vspace{0.3cm}\noindent
Based on the comments of both reviewers and my own careful consideration, I unfortunately must reject this manuscript.

\vspace{0.3cm}\noindent
If you choose to address the problems identified by the reviewers and resubmit your manuscript, please upload it to the Editorial Manager as a new submission and follow the AMS Resubmission Requirements at: www.ametsoc.org/PubsResubmission. You will need to declare it as a resubmission and include a Response to Reviewers.

\vspace{0.3cm}\noindent
Along with your substantially revised manuscript, please upload a point-by-point response that satisfactorily addresses the concerns and suggestions of each reviewer and the Editor. To help us to assess your revisions, our journal recommends that you cut-and-paste the reviewer and Editor comments into a new document. As you would conduct a dialog with someone else, insert your responses in a different font, font style, or color after each comment. If you have made a change to the manuscript, please indicate where in the manuscript the change has been made. (Indicating the line number where the change has been made would be one way, but is not the only way.) Although it is not required by our journal, you may wish to include a tracked-changes version of your manuscript. You will be able to upload this as "additional material for reviewer reference." Should you disagree with any of the proposed revisions, you will have the opportunity to explain your rationale in your
response. No separate cover letter to me is needed unless it contains essential information that does not appear in your reply.

\vspace{0.3cm}\noindent
Thank you for your interest in Journal of the Atmospheric Sciences. Please contact me at Stern.JAS@ametsoc.org if you have any questions. We wish you a more positive result with future endeavors.

\vspace{0.3cm}\noindent
Best regards,

\vspace{0.3cm}\noindent
Daniel Stern

\noindent
Editor

\noindent
Journal of the Atmospheric Sciences

\vspace{0.5cm}
{\color{blue}\normalfont
\noindent
Date: 24 January 2026

\vspace{0.3cm}\noindent
To:

\noindent
Dr. Daniel Stern, Handling Editor

\noindent
Dr. Zhuo Wang, Editor in Chief

\noindent
Journal of the Atmospheric Sciences

\vspace{0.3cm}\noindent
Subject: Editorial action regarding Sparks and Toumi (2022)

\vspace{0.3cm}\noindent
Dear Dr. Stern and Dr. Wang,

\vspace{0.3cm}\noindent
We write concerning our Comment on Sparks and Toumi (2022), which identifies an unjustified omission of the precipitation term from the pressure-tendency equation. We are grateful to the reviewers for their careful evaluation.

\vspace{0.3cm}\noindent
One reviewer stated explicitly that the criticism is valid. A second reviewer characterized it as potentially valid but did not address its substance, instead suggesting development of a standalone paper with a new theory. A third reviewer noted that the model of Sparks and Toumi (2022) would be valid if total air density were replaced by dry-air density, acknowledging that this should be noted. Importantly, none of the reviewers refuted our central claim that, as currently formulated, the model is invalid.

\vspace{0.3cm}\noindent
While we respect the editorial decision and recognize that judgments of scope and format may differ, we are concerned that the present outcome leaves an acknowledged inconsistency unaddressed. In its published form, the model does not specify dry-air density in the governing equations.

\vspace{0.3cm}\noindent
Although total air density and dry-air density differ by only about one percent, the key diagnostic quantity—the column-mean velocity—can differ by more than 100\% and even change sign depending on which density is used. This is therefore not a minor issue. Moreover, while the equations could in principle be reinterpreted, the numerical analyses cannot be retroactively corrected: they necessarily used either total air density or dry-air density, and this choice must be made explicit.

\vspace{0.3cm}\noindent
In our view, if total air density was used, publication of our Comment would be the appropriate mechanism to clarify the issue for readers. If dry-air density was used, a corrigendum would be required. In either case, leaving the matter unresolved does not seem consistent with AMS standards of scientific rigor and self-correction.

\vspace{0.3cm}\noindent
We therefore respectfully ask that the journal either reconsider publication of our revised Comment, which addresses all reviewer comments, or request that the authors of Sparks and Toumi (2022) publish a corrigendum clarifying the formulation.

\vspace{0.3cm}\noindent
Thank you for your time and consideration. We look forward to your response.

\vspace{0.3cm}\noindent
Yours sincerely,

\noindent
Anastassia Makarieva
}

}


\end{document}